\def\Den{\hbox{Den}}
\def\den{\hbox{den}}
\def\hc{\hbox{h.c.}}
\def\tr{\hbox{tr}}
\def\ln{\ell{n}}
\documentstyle[12pt]{article}

 {\parindent 0.5cm \textwidth 15.0cm \textheight
23.5cm \hoffset=-0.8cm \voffset=-3cm
  \let\LARGE=\large
 \let\large=\normalsize
\newcommand{\be}{\begin{equation}}
\newcommand{\ee}{\end{equation}}
\newcommand{\ba}{\begin{array}{c}}
\newcommand{\ea}{\end{array}}

\begin{document}
\begin{titlepage} \vspace{0.2in} \begin{flushright}
MITH-94/11 \\ \end{flushright} \vspace*{1.5cm}
\begin{center} {\LARGE \bf  A Possible Origin of the Wilson Lattice Fermion
\\} \vspace*{0.8cm}
{\bf Giuliano Preparata and She-Sheng Xue$^{a)}$}\\ \vspace*{1cm}
INFN - Section of Milan, Via Celoria 16, Milan, Italy\\ \vspace*{1.8cm}
{\bf   Abstract  \\ } \end{center} \indent

Considering relevant and irrelevant high-dimension operators for ordinary and
mirror fermions respectively, we show that the Wilson lattice fermion can
originate from a spontaneous symmetry breaking phenomenon. Ward identities, due
to symmetries at the cutoff, guarantee the cancellation of mirror-fermion
contributions and the infrared limit can be achieved. The Wilson parameter
turns out to be fixed at $r\simeq 0.18$, where the vacuum energy of the system
is minimized.
\vfill \begin{flushleft}  April, 1994 \\
PACS 11.15Ha, 11.30.Rd, 11.30.Qc  \vspace*{3cm} \\
\noindent{\rule[-.3cm]{5cm}{.02cm}} \\
\vspace*{0.2cm} \hspace*{0.5cm} ${}^{a)}$
E-mail address: xue@milano.infn.it\end{flushleft} \end{titlepage}

\noindent
{\bf 1.}\hspace*{0.3cm} The fermion ``doubling'' phenomenon is a well-known
problem arising when fermion fields are defined on a lattice. In fact, this
spectrum doubling is quite a general property of regularized fermion theories
\cite{nogo}.  However, this ``problematic'' phenomenon, instead of being a
shortcoming of any reasonable regularization, perhaps gives us a physical hint
to understand the short-distance behaviour of the Standard Model.
Since chiral symmetry is perfectly conserved, the
space-time lattice provides a very well-defined regularization for a fermionic
field theory. In order for this occurrence, mirror fermions participate
dynamically
in the spectrum of the theory. In contrast, from a dynamical viewpoint,
we have not completely understood the operator content
of the theory at short distances, where there must exist local and
non-local high-dimension operators for ordinary and mirror fermions and gauge
bosons, owing to the experimental observation of a rich mass spectrum of
fundamental particles in the low-energy region. At long distances, these
high-dimension operators, however, should be relevant and irrelevant for
ordinary fermions and mirror fermions respectively in such a way that mirror
fermions decouple from the low-energy spectrum and ordinary fermions remain and
couple properly with gauge bosons.

Let us consider lattice QCD as an example. It has been suggested that lattice
regularization, avoiding fermion doubling, can be achieved
by introducing
an extra dimension-5 operator (the Wilson term) into the naive lattice
fermion action $S_d(U,\bar\psi,\psi),\;(U\in SU(N_c))$ \cite{wilson},
\begin{equation}
S=S_d(U,\bar\psi,\psi)+{r\over a}\sum_{x,\mu}\bar\psi(x)\partial_\mu^2\psi(x),
\label{wilson}
\end{equation}
where $a$ is the lattice spacing and the lattice ``laplacian'' $\partial_\mu^2$
is defined as
\begin{equation}
\partial_\mu^2\psi(x) = U_\mu(x)\psi(x+a_\mu)+
U^\dagger_\mu(x)\psi(x-a_\mu)-2\psi(x).
\label{laplacian}
\end{equation}
This extra term with a finite Wilson parameter $r$ ($0 < r \leq 1$ for
reflection positivity) becomes an irrelevant operator for ordinary fermions in
the infrared limit, while it is a relevant operator for mirror fermions in the
high-energy regime, in fact it generates an effective mass $M\sim {r\over a}$
for mirror fermions. If we wish to achieve a chiral-invariant QCD in the
continuum limit, we must carefully tune {\it only} one free parameter in our
lattice action so  that intrinsic quark mass $m$ goes to zero as the continuum
limit is approached. However, at the same time, the axial anomaly of the theory
is restored \cite{smit} without any Goldstone modes, due to the explicit
breaking of chiral symmetry in the Wilson term. In this letter, we would like
to demonstrate that the Wilson fermion can be dynamically generated from
chiral-invariant
Nambu-Jona Lasinio (NJL) interactions \cite{nambu} of ordinary and mirror
fermions.

\vspace*{0.5cm}
\noindent
{\bf 2.}\hspace*{0.3cm}
This attempt leads us to consider the following lagrangian,
\begin{eqnarray}
S&=&S_g(U)+S_d(\bar\psi,\psi,U)+S_r+S_{ir},\label{action}\\
S_r &=&-G_1\sum_x\bar\psi_L(x)\cdot\psi_R(x)
\bar\psi_R(x)\cdot\psi_L(x),\label{sr}\\
S_{ir} &=&-{G_2\over 2}\sum_{\mu,x}\bar\psi_L(x)\cdot\partial^2_\mu\psi_R(x)
\bar\psi_R(x)\cdot\partial^2_\mu\psi_L(x),\label{sir}
\end{eqnarray}
where $S_g$ is the usual Wilson gauge-action and $G_{1,2}$ are two,
as yet unspecified,
Fermi-type $O(a^2)$ coupling constants. Note that (i) in the naive continuum
limit ($S_{ir}\simeq 0$), eq.~(\ref{action}) is
just a gauged NJL model for ordinary fermions
($pa\ll 1$)\cite{bar}; (ii) in the naive
``lattice limit'' $S_{ir}$ is a NJL interaction for mirror fermions
($pa\simeq 1$) and $S_r$ is a NJL interaction in which both ordinary
and mirror fermions participate; (iii)
obviously, $S_{r} (S_{ir})$ is a relevant (irrelevant) operator for ordinary
fermions and both $S_{r}, S_{ir}$ are relevant operators for mirror fermions.

Using the Landau mean-field (tree-level) and
the large-$N_c$ approach ($N_c\gg 1, N_cG_{1,2}$ fixed),
we consider the leading $O(N_c)$ terms of the gap equation for the fermion
self-energy function $\Sigma(ap)$:
\begin{equation}
\Sigma(ap) =  2g_1 \int_l {\Sigma(l)
 \over \Den(l)} + {g_2\over 2} \int_l w(l) {\Sigma(l) \over \Den(l)}w(ap)
+\Sigma_g(ap),
\label{gap0}
\end{equation}
where $g_{1,2} a^2 = N_c G_{1,2}; l_\mu = q_\mu a,
\int_l = \int^\pi _{- \pi} {d^4l
\over (2\pi)^4}; w(l)=\sum_\mu(1-\cos l_\mu)$ and
$\Den(l) = \sin^2l_\mu + (a\Sigma(l))^2$. $\Sigma_g(ap)$ is the self-energy
function due to the gauge interaction.
On the basis of the structure of gap equation
(\ref{gap0}) and $\Sigma_g(ap)$ \cite{smit}
one may rewrite $\Sigma(ap)=\Sigma'(ap)+\bar rw(ap),\;(r=a\bar r)$ and
for arbitrary external momenta $p$, eq.~(\ref{gap0})
is equivalent to
\begin{eqnarray}
\Sigma'(ap) &=& 2g_1 \int_l {\Sigma'(l) +\bar rw(l)
\over \Den(l)}+\Sigma_g(ap);\label{gap1}\\
\bar r &=& {g_2\over 2} \int_l w(l) {\Sigma'(l) + \bar rw(l) \over \Den(l)},
\label{gap2}
\end{eqnarray}
where $\bar r$ is the following condensate of mirror fermions in the mean-field
and large-$N_c$ approximation,
\begin{equation}
\bar r= {G_2\over 4}\!\sum_{\mu,x}\left({1\over 4}\right )
\left\langle\bar\psi_L(x)\partial_\mu^2\psi_R(x)+\hc\right\rangle
.\label{r}
\end{equation}
We clearly find that $\bar r$ stems from the contribution of mirror fermions
owing to the factor $w(l)$ in eq.~(\ref{gap2}) and $\Sigma'(ap)$ in
eq.~(\ref{gap1}) received contributions from both ordinary and mirror fermions.

For small external momenta $pa\ll 1$ in eq.~(\ref{gap0}), we introduce an
arbitrary scale $\epsilon\; (pa\ll\epsilon\ll\pi)$ separating the ``lattice
region'' $(\epsilon,\pi)^4$, populated by mirror fermions, from the
``continuum region''$(0,\epsilon)^4$, where ordinary fermions lie. For
small internal momenta $l\in (0,\epsilon)^4,\;w(l) \simeq 0$ and $\bar
r\rightarrow 0^+$ (see eq.~(\ref{gap2})), the gap equation (\ref{gap0}) should
reduce to its counterpart in the continuum theory. The mirror-fermion
contributions to the gap-equation (\ref{gap0}) come from the ``lattice
region'', where internal momenta are large $l\in (\epsilon,\pi)$. Based on this
separation, we write eq.~(\ref{gap1}) for $pa\ll 1$ as
\begin{equation}
\Sigma(ap) =  2g_1 \int_{l\in (0,\epsilon)} {\Sigma'(l)
 \over \Den(l)}+2g_1 \int_{l\in (\epsilon,\pi)} {\Sigma'(l)
\over \Den(l)}+2g_1 \int_{l\in (\epsilon,\pi)}
{\bar r w(l) \over \Den(l)} +\Sigma_g(ap),
\label{ep}
\end{equation}
where we have $\Sigma_g(ap)
=\Sigma^c_g(ap)+{\delta(r)\over a}$ \cite{smit,xue}, where $\Sigma^c_g(ap)$ is
the self-energy function $\Sigma_g(ap)$ in the ``continuum region'' and
$\delta(r)$ is a finite function of $r$.
The limit $\epsilon\rightarrow 0$ can be taken in the third term owing to the
factor $w(l)$.  The $\epsilon$-dependence
of the first term must be cancelled by that of the
second term in eq.~(\ref{ep}).
The value of $\epsilon$ is chosen so that in $\Sigma'(l)$ we separate the
$O({1\over a})$
divergent part from the finite part $\Sigma_c(l)\;(a\Sigma_c(l)\ll 1)$, which
is the counterpart of $\Sigma(l)$ in the continuum theory. One can thus rewrite
eq.~(\ref{ep}) as
\begin{equation}
\Sigma_c(ap) =  2g_1 \int_l {\Sigma_c(l)
 \over \Den(l)}+{\Delta\over a}+2g_1 \int_l {
\bar r w(l) \over \Den(l)} +\Sigma^c_g(ap)+{\delta(r)\over a},
\label{de}
\end{equation}
where
\begin{equation}
{\Delta\over a}\simeq 2g_1 \int_{l\in (\epsilon,\pi)} {\Sigma'(l)
\over \Den(l)},
\label{mc}
\end{equation}
which is a mirror-fermion contribution via the NJL interaction $S_r$
in (\ref{action}).

The symmetry of the action (\ref{action}) implies Ward identities
satisfied by Green functions on
the lattice. These Ward identities must be obeyed up to powers of the
lattice spacing $O(a)$. The Ward identity due to the chiral symmetry of the
action
(\ref{action}) at the cutoff
\begin{equation}
\langle\psi_L(0)\bar\psi_R(x)\rangle
=0\hskip2cm (x\gg a)
\label{w}
\end{equation}
implies that $O({1\over a})$ divergences in eq.~(\ref{de}) should cancel
\begin{eqnarray}
\Sigma_c(ap) &=&  2g_1 \int_l {\Sigma_c(l)
 \over \den(l)}+\Sigma^c_g(ap)\label{cgp}\\
0&=&{\Delta\over a}+{2g_1\over a} \int_l {
r w(l) \over \den(l)} +{\delta(r)\over a},
\label{t}
\end{eqnarray}
where it is self-consistent that $\den(l)=\sin^2(l)+(\Sigma_c(l)a+rw(l))^2$.
We find that ${\Delta\over a}$ acts as the mass counterterm considered
in the Rome approach \cite{rome}, where the Ward identities are due to the
BRST symmetry. Eq.~(\ref{cgp}) is analogous to the gap equation in the
continuum theory
\cite{bar} containing gauge and four-fermion interactions.
In fact, the infrared scale $m=\Sigma_c(0)$, i.e., the
v.e.v.~of the Standard Model, and the mass counterterm
${\Delta\over a}$ are contributions of ordinary fermions and mirror fermions
respectively to the following condensate,
\begin{equation}
M =-{G_1\over 2}\sum_x\langle\bar\psi(x)
\psi(x)\rangle\simeq m+{\Delta\over a}.\label{mass}
\end{equation}
It is clear that at short distances (the cutoff), the theory should recover
the symmetries of action (\ref{action}). This requires that all hard breaking
terms $O({1\over a})$, i.e., mirror-fermion contributions should cancel
at the cutoff. Such cancellation is guaranteed by the Ward identities.

\vspace*{0.5cm}
\noindent
{\bf 3.}\hspace*{0.3cm}
We now turn to the discussion of whether there exists a consistent solution of
eqs.~(\ref{gap2},\ref{cgp},\ref{t}) in the infrared limit
($\Sigma_c(ap)a\ll 1$). Since $\Sigma_c(ap)a\simeq 0$, eq.~(\ref{gap2})
approximately implies
\begin{equation}
1 = {g_2\over 2} \int_l w(l) {w(l) \over \sin^2 l_\mu + (r)^2 w(l)^2},
\label{rg}
\end{equation}
establishing a relationship between $g_2$ and $r$,
which is reported in Fig.(1). As one can see, we find that the coupling $g_2$
has to be strong enough ($g_2 > g^c_2\simeq 0.2$) in order for $r\sim O(1)$.
Eq.~(\ref{cgp}) can be solved analytically. As discussed
in refs.~\cite{bar1,xue}, a critical line, where the infrared limit (
$a\Sigma_c(ap)\ll 1$) can be defined,
is found in terms of $g_1$ and gauge coupling. The Ward identity (\ref{w}) thus
determines the mass counterterm ${\Delta\over a}$ as a function of $g_1$ and
gauge coupling.

Composite modes are bound to be produced once the spontaneous
symmetry breakdown occurs $m\not=0$, as evidenced by the non-trivial solutions
of the gap equations (\ref{cgp}). In order to see these modes, we calculate the
four-fermion scattering amplitudes associated with the vertex $S_r$ within the
approximation of $\Sigma_c(ap)\simeq m$ and neglecting gauge field. This
calculation is straightforward and we just present the results. Composite modes
in the pseudo-scalar channel $\Gamma_p(q^2)$ and the scalar channel
$\Gamma_s(q^2)$ are
\begin{eqnarray}
\Gamma_p(q^2) &=& {{1\over2}} {1 \over I_p(q) A(q^2)},\label{gold}\\
\Gamma_s(q^2) &=& {{1\over2}} {1 \over {4N_c \over a^2} \int_k {[ma +rw(k)]^2
\over
D(k,qa)} + I_s(q)A(q^2)}\label{scalar},
\end{eqnarray}
where
\begin{eqnarray}
A(q^2) &=& \sum_\mu \left({2 \over a} \sin {q_\mu a \over 2}\right)^2,
\nonumber\\
I_p(q) &=& {N_c \over 4} \int_k {c^2 (k) + (r)^2 s^2(k) \over D(k,qa)},
\nonumber\\
I_s(q) &=& {N_c\over 4} \int_k {c^2(k) \over D(k,qa)}\label{i},
\end{eqnarray}
where $c^2(k) = \sum_\mu \cos^2 k_\mu, s^2(k) = \sum_\mu \sin^2 k_\mu$ and
$D(k,qa)=\den (k+{qa\over 2})\den (k-{qa\over 2})$. We find a massless
Goldstone mode, which should be a candidate for the longitudinal mode of
the massive
gauge boson, and a scalar mode, which should be a candidate for the Higgs
particle. The scalar mode disappears from the
low-energy spectrum since its mass is proportional to ${r\over a}$, a term
which is contributed by mirror fermions.

\vspace*{0.5cm}
\noindent
{\bf 4.}\hspace*{0.3cm}
In order to ascertain the stability, and hence the physical realizability, of
the solution $\Sigma_c(ap)\not=0$ and $r\not=0$ to eqs.~(\ref{gap2},\ref{cgp}),
we turn to the computation of the ground
state energy. In the one-loop approximation ($O(N_c)$), $\Sigma_c(ap)\simeq m$
and neglecting the gauge field, the effective potential
is given by
\begin{equation}
V(M,\bar r)={M^2\over G_1}+
{\bar r^2\over G_2}-N_c\tr\int_l\ln
\{ {\gamma_\mu\sin l_\mu\over a}+(M+\bar rw(l))\}+\cdot\cdot\cdot,
\label{eff}
\end{equation}
and the difference between the energy
of the symmetric vacuum and broken vacuum $\Delta E_\circ=V(M,r)-V(0,0)$ is
given by
\begin{equation}
\Delta E_\circ= -{2N_c \over a^4} \int_l \sum^{\infty} _{k=1}
{1 \over k+1}
\left[{(ma + rw(l))^2 \over s^2(l) + (ma + r w(l))^2} \right]^{k+1},
\label{vac1}
\end{equation}
which is obtained from (\ref{eff}) by considering the gap equation
(\ref{gap1},\ref{gap2}) $(M\simeq\Sigma(l))$. This shows that the non-trivial
solutions of the gap
equations characterize a chirally asymmetric vacuum that has an energy density
lower than that of the symmetric vacuum. However, $r$ gets the largest value
permitted by eq.~(\ref{gap2}). On the other hand, noticing that composite
bosons
give a positive energy density of such broken ``vacua'' and this positive
contribution certainly increases as $r^2$ increases, we turn to the
computation of the total vacuum energy ($\Delta E$) containing both fermion and
composite
boson contributions on the basis of the one-loop gap equations,
(\ref{gap2},\ref{cgp}),
\begin{equation}
\Delta E = -\left[\ln\int_f\exp(-S_{eff}(m,r))-\ln\int_f\exp(-S_{eff}(0,0))
\right],
\label{tol}
\end{equation}
where $S_{eff}(m,r)$ is the effective Wilson action over the ground state.
The details of the calculation are lengthy and will not be reported in this
letter, we just present the result( $\Delta E_\circ$ is $O(N_c)$ and the
second term $O(N_c^0)$):
\begin{eqnarray}
\Delta E &\simeq& \Delta E_\circ-\Big[1-e^{-{\Delta E_\circ\over
N_c}}\Big]\nonumber\\
&&\cdot\left[-4+\int_l\big(4g_1\tilde\Gamma_s(l)
+{1\over4g_1\tilde\Gamma_s(l)}\big)\!+\!
\int_l\big(4g_1\tilde\Gamma_p(l)+{1\over4g_1\tilde\Gamma_p(l)}
\big)\right],
\label{vac2}
\end{eqnarray}
where
\begin{eqnarray}
\tilde\Gamma_p(l)&=&{a^2I_p(l)A(l^2)\over 4N_c};\nonumber\\
\tilde\Gamma_s(l)&=&{a^2\over 4N_c}\left({4N_c \over a^2} \int_k
{[ma +rw(k)]^2 \over
D(k,l)} + I_s(l)A(l^2)\right).
\label{extra}
\end{eqnarray}
Combining positive and negative contributions in eq.~(\ref{vac2})
and putting $ma\simeq 0$, $N_c=3$ and $g_1\simeq g_1^c(r)$ obtained from
eq.~(\ref{cgp}) with $\Sigma^c_g(ap)\simeq 0$, we find (Fig.(2)) that
$r\simeq 0.18$ is the energetically favoured solution. This shows that the
effective Wilson action is stable over the chiral-symmetry-violating
ground state.

\vspace*{0.5cm}
\noindent
{\bf 5.}\hspace*{0.3cm}
The high-dimension operators $S_r$ and $S_{ir}$ at the cutoff (\ref{action})
can probably be induced from the quantum gravity \cite{xp}. These operators
evade {\it
in principle} the ``no-go'' theorem. We have shown by considering the Dyson
equation for the fermion sector and the vacuum energy that the operators $S_r$
and
$S_{ir}$ can produce the low-energy spectrum $\Sigma_c(ap)$, the consistent
mass
counterterm ${\Delta\over a}$ and the Wilson parameter $r$ that removes mirror
fermions. Owing to the vector-like gauge symmetry of (\ref{action}), the gauge
field does not couple to the composite modes and the vacuum polarization is
transverse and free from the $O({1\over a^2})$ divergence. Logarithmic
divergences
in $\Sigma_c(ap)$, vacuum polarization and vertex function are treated in the
normal renormalization prescription. As the ordinary-fermion mass $m$ is tuned
to be zero, the axial anomaly is reproduced by $r\not=0$ \cite{smit} without
any
extra massless Goldstone modes \cite{xg}. Thus in the context of a vectorial
gauge theory with extended high-dimension operators (\ref{action}), the
``no-go'' theorem is evaded also {\it in practice}.

As for a chiral gauge theory, it should be mentioned that the E-P model
\cite{ep}, which carefully breaks unwanted global symmetries, preserves the
chiral gauge symmetries and gets rid of mirror fermions through the
perturbation of four-fermi interactions. It was claimed \cite{pg} that the E-P
model fails to reach its goal on the basis that the E-P model and the
Smit-Swift model \cite{ss} are in the same universality class and spontaneous
breaking of symmetry occurs. This ``universality''
\cite{ku} between the ``$\bar tt$-condensate model'' \cite{bar2} and the Higgs-
Yukawa model was shown based on renormalization group
arguments (which however have been recently questioned \cite{ra} in the strong
coupling
region). According to the spirit of the approach
presented in this paper, the key points for a possible lattice chiral gauge
theory
are: (i) whether there exist high-dimension operators ($S_r, S_{ir}$ {\it
etc.}) that preserve chiral gauge symmetries at the cutoff, generating all
necessary counterterms as carefully discussed in the Rome approach \cite{rome};
(ii) whether Ward identities stemming from symmetries at the cutoff (instead of
the BRST symmetry implemented in the Rome approach) could consistently
guarantee
the achievement of a sensible low-energy chiral gauge theory. We are left
with this problem for future work.

\newpage  \pagestyle{empty}
\begin{center} \section*{Figure Captions} \end{center}

\vspace*{1cm}

\noindent {\bf Figure 1}: \hspace*{0.5cm}
 The function $r(g_2)$ in terms of $g_2$ ($g_2 > g_2^c\simeq 0.2$).

\noindent {\bf Figure 2}: \hspace*{0.5cm}
 The vacuum energy $\Delta E(r)$ in terms of the Wilson parameter $r$
($r_m\simeq 0.18$).

}
\end{document}